# Mechanism of Quercetin in Inhibiting Triple-Negative Breast Cancer by Regulating T Cell-Related Targets: An Analysis Based on Single-Cell Sequencing and Network Pharmacology

Chen Ruiqi[1], Han Liang[1,2,&], Wang Fengyun[1,&]

1. College of Traditional Chinese Medicine, Guangdong Pharmaceutical University, Guangzhou 510006

2. School of Health, Guangdong Pharmaceutical University, Guangzhou 510006

&Corresponding author Email:hanliang72@163.com, wfycn2000@163.com

**Abstract**

Objective: To investigate the mechanism by which quercetin inhibits triple-negative breast cancer (TNBC) through regulating T-cell-related targets, providing a novel strategy for TNBC immunotherapy.Methods: Single-cell RNA sequencing (GSE161529 dataset) and network pharmacology were integrated. PCA and UMAP clustering identified T-cell subsets and differentially expressed genes in TNBC microenvironment. TNBC-related targets were screened via CTD and OMIM databases, with functional pathways analyzed by GO/KEGG enrichment. Molecular docking and PPI networks validated interactions between quercetin and core targets.Results: Quercetin intersected with 79 TNBC targets, including AKT1, EGFR, and MMP9, enriched in EGFR inhibitor resistance and endocrine resistance pathways. Molecular docking revealed the highest affinity between quercetin and GSK3B (-13.2 kJ/mol). AKT1 and MMP9 expression correlated with patient survival.Conclusion: Quercetin may reverse TNBC immunosuppression by multi-target modulation of T-cell function, but clinical application requires solutions for its low bioavailability, such as delivery systems or combination therapies.

# 1. Introduction

Triple-negative breast cancer (TNBC) is a highly aggressive and poor-prognosis subtype of breast cancer, accounting for approximately 15% to 20% of all breast cancer cases [1]. Its pathological features are characterized by the absence of expression of estrogen receptor (ER), progesterone receptor (PR), and human epidermal growth factor receptor 2 (HER2), thus making it unresponsive to endocrine therapy (such as tamoxifen) or HER2-targeted therapy (such as trastuzumab) [2]. Currently, chemotherapy remains the main treatment modality for TNBC. However, due to its high heterogeneity, early metastasis, and chemotherapy resistance, the 5-year survival rate of TNBC patients is significantly lower than that of other breast cancer subtypes [3]. Moreover, the tumor microenvironment (TME) of TNBC has strong immunosuppressive properties, further complicating treatment challenges [4]. In recent years, immune checkpoint inhibitors (such as PD-1/PD-L1 antibodies) have shown some potential in the treatment of TNBC, especially for PD-L1 positive patients. However, the clinical response rate is still limited, and some patients develop resistance or disease progression quickly [5]. Studies have shown that T cells play a key role in the immune microenvironment of TNBC, and their functional status directly affects the tumor's immune escape ability [6]. For instance, the activity of CD8+ T cells in tumor-infiltrating lymphocytes (TILs) is positively correlated with the prognosis of TNBC patients [7], while the enrichment of regulatory T cells (Tregs) may promote immunosuppression [8]. Despite this, the specific molecular mechanisms underlying T cell dysfunction in TNBC have not been fully elucidated, and which key targets can be modulated by drugs to enhance anti-tumor immune responses still requires further exploration.

Quercetin (3,3',4',5,7-pentahydroxyflavone) is a natural flavonoid compound widely found in fruits (such as apples, citrus fruits), vegetables (such as onions, broccoli), and traditional Chinese medicines (such as Ginkgo biloba leaves, mulberry leaves) [9]. The multiple phenolic hydroxyl groups in its chemical structure endow it with strong antioxidant activity, which can reduce oxidative stress damage by scavenging free radicals and chelating metal ions (such as $Fe^{2+}$, $Cu^{2+}$) [10]. Additionally, quercetin has significant anti-inflammatory effects, being able to inhibit the expression of pro-inflammatory factors such as NF-κB, COX-2, and iNOS, thereby regulating the chronic inflammatory microenvironment [11].

In the field of anti-tumor research, quercetin has been proven to inhibit tumor occurrence and development through multiple pathways: 1. Direct action on tumor cells: by regulating the cell cycle to induce G1 phase arrest [12], or activating the mitochondrial apoptotic pathway to promote tumor cell apoptosis [13]. 2. Inhibiting metastasis and angiogenesis: quercetin can reduce the activity of MMP-2/9 to decrease extracellular matrix degradation, and simultaneously inhibit the VEGF signaling pathway, thereby hindering tumor invasion and angiogenesis [14]. 3. Regulating epigenetic modifications: recent studies have found that quercetin can reverse the silencing of tumor-related genes by inhibiting DNA methyltransferases (DNMTs) and histone deacetylases (HDACs) [15].

Notably, the regulatory effect of quercetin on the tumor immune microenvironment has gradually become a research hotspot. Preclinical studies have shown that quercetin can enhance anti-tumor immune responses through the following mechanisms: 1. Promoting T cell activation: upregulating the expression of co-stimulatory molecules on the surface of CD4+ and CD8+ T cells, enhancing their proliferation and cytotoxic functions. 2. Reversing immunosuppression: inhibiting the immunosuppressive activity of Treg cells and reducing the recruitment of myeloid-derived suppressor cells (MDSCs). 3. Regulating the cytokine network: Quercetin reduces the secretion of immunosuppressive factors such as TGF-β and IL-10, while promoting the release of pro-inflammatory factors such as IFN-γ and TNF-α [16]. However, there are still key scientific questions regarding how quercetin precisely regulates T cell function in the special subtype of TNBC: 1. Insufficient target specificity: The action targets of quercetin involve multiple proteins such as kinases (e.g., PI3K, EGFR), transcription factors (e.g., STAT3), and epigenetic enzymes, but its direct action targets in TNBC T cells have not been systematically identified. 2. Limited mechanism research: Existing studies mostly focus on the effects of quercetin on tumor cells, while the regulatory mechanism of quercetin on the TNBC immune microenvironment (especially the interaction network of T cell subsets) lacks single-cell level analysis. 3. Bottlenecks in translational application: The low bioavailability of quercetin may limit its clinical efficacy, and it is necessary to combine delivery systems or structural optimization strategies to improve its targeting.

This study, based on single-cell RNA sequencing data from the GEO database

(GSE161529), identified T cell subsets and their differentially expressed genes in the TNBC tumor microenvironment through principal component analysis (PCA) and UMAP clustering techniques. Further, TNBC-related targets were screened by integrating databases such as CTD, OMIM, and DisGeNET, and their functional pathways were revealed through GO/KEGG enrichment analysis. Using network pharmacology methods, Chinese medicines and active components related to key TNBC targets were screened from SymMap and TCMSP databases, and quercetin was identified as a potential therapeutic molecule. Molecular docking and protein-protein interaction (PPI) network analysis were used to verify the binding ability of quercetin to core TNBC targets (such as AKT1, EGFR, MMP9, etc.) and its impact on patient prognosis.

This study aims to clarify the molecular mechanism by which quercetin inhibits TNBC through the regulation of T cell-related targets, providing new theoretical basis and potential drug candidates for the immunotherapy of TNBC.

## 2. Methods

2.1 Data source

Single-cell sequencing data related to "triple negative breast cancer" (TNBC) were retrieved from the GEO database (https://www.ncbi.nlm.nih.gov/geo/). The 10X Genomics Chromium data of TNBC (GSE161529) was selected as the data source, which included a total of 421,761 cells from 52 patients. This included profiles of 4 TNBC cases, 4 BRCA1 TNBC cases, 6 HER+ tumors, 19 ER+ tumors, and 6 ER+ tumor lymph node metastases. It also included the profiles of total breast cells from 13 normal breast cancer-free patients, epithelial breast cells from 11 normal patients, and total breast cells from 4 BRCA1 mutation pre-cancer patients. TNBC patients and normal samples were selected as the research data [17].

2.2 Principal Component Analysis (PCA) and UMAP, tSNE Clustering Analysis

The single-cell transcriptome sequencing data of each group were analyzed using the Seurat package in R language software. Firstly, the "limma" package in R software was used to remove genes expressed in less than 3 cells and cells expressing less than 200 genes, and only cells with 200-6000 genes detected were retained as the threshold for identifying high-quality single-cell sequencing samples. To narrow the range of gene screening, linear dimensionality reduction and visualization were performed through PCA analysis. PCA was conducted on the single-cell data set to calculate the main components of dimensionality reduction, and an appropriate value was selected for the next step of dimensionality reduction analysis. UMAP (Uniform Manifold

Approximation and Projection) is a nonlinear manifold learning technique for dimensionality reduction. It aims to map high-dimensional data to a low-dimensional space (usually two or three dimensions) for visualization or data compression while preserving the global structure and local neighborhood relationships of the data as much as possible. tSNE (t-distributed stochastic neighbor embedding) is an unsupervised learning method for dimensionality reduction, which can effectively reduce the dimension of data and visually represent the clustering clusters of different cell populations. Cell types were annotated by referring to the human cell marker genes in the CellMarker database or the transcriptome/markers of related cells in published literature for subsequent research.

2.3 Collection of Target Points Related to Triple-Negative Breast Cancer

In databases such as CTD (http://ctdbase.org/), OMIM (https://www.omim.org/), PharmGkb (https://www.pharmgkb.org/), DisGeNET (https://www.disgenet.org/), and GeneCards (https://www.genecards.org/), target points related to triple-negative breast cancer were screened using the keyword "triple negative breast cancer". Duplicates were removed, and then the intersection was taken with the differentially expressed genes of T cells obtained from GEO. A Venn diagram was drawn.

2.4 Enrichment of Intersection Targets between T-cell Differential Genes and Triple-negative Breast Cancer

The intersection targets between T-cell differential genes and triple-negative breast cancer were subjected to GO (Gene Ontology) enrichment analysis and KEGG (Kyoto Encyclopedia of Genes and Genomes) enrichment analysis through the Bioinformatics website (https://www.bioinformatics.com.cn/).

2.5 Identify Key Genes Based on Enriched Pathways

Search for the enriched pathways in Genecard (https://www.genecards.org/), and find the top 50 genes with a score greater than 60 in each pathway. Calculate the frequency of the related genes and determine the top 10 genes in this ranking as the key genes.

2.6 Screening of Traditional Chinese Medicines Related to Triple-Negative Breast Cancer

Input these key genes in SymMap (http://www.symmap.org/) and screen the related traditional Chinese medicines under the conditions of P_value < 0.05, FDR (BH) < 0.05, and FDR (Bonferroni) < 0.05. Then, conduct statistical analysis on the meridian tropism and taste of the related traditional Chinese medicines.

2.7 Collection of Active Components of Relevant Traditional Chinese Medicines and Their Target Sites

The chemical components of the relevant traditional Chinese medicines were retrieved and screened from the Traditional Chinese Medicine Systems Pharmacology Database and Analysis Platform (TCMSP, https://tcmspe.com/). The active component groups of the relevant traditional Chinese medicines were screened under the conditions of oral bioavailability (OB) ≥ 30% and drug-likeness (DL) ≥ 0.18. The target site information of the active components was obtained from this database. The relevant active components were statistically analyzed, and the most frequently occurring active components were selected as the research objects to explore whether they could target T cells for anti-tumor immunity.

2.8 Key Targets of Quercetin in Triple-Negative Breast Cancer

The SMILES number of quercetin was retrieved from Pubchem (https://pubchem.ncbi.nlm.nih.gov/) and imported into Swiss Target Prediction (http://swisstargetprediction.ch/) to predict its target proteins. The intersection of the targets of quercetin and those related to triple-negative breast cancer was taken to obtain the key targets of quercetin in triple-negative breast cancer.

2.9 Enrichment Analysis of Quercetin Targets in Triple-Negative Breast Cancer

Through the MicroBioinfo website (https://www.bioinformatics.com.cn/), KEGG (Kyoto Encyclopedia of Genes and Genomes) enrichment analysis was conducted on the targets of quercetin in triple-negative breast cancer. The conditions were set as follows: species defined as "Homo sapiens", P ≥ 0.05, to determine the core pathways that may be involved.

2.10 Construction of Quercetin and Triple-Negative Breast Cancer Protein-Protein Interaction (PPI) Network and Core Target Analysis

The STRING database (https://string-db.org/) is a database that establishes direct and indirect interactions between known and predicted proteins, and can visually present the relationship network between protein structures and functions. The key target genes of quercetin intervention in triple-negative breast cancer were imported into the STRING database, and the research species was selected as human (Homo sapiens) to obtain the PPI network. Cytoscape software was used for visualization analysis and calculation of the degree value. The top 10 core genes (Degree > 20) were selected as core targets.

2.11 Expression Analysis of Core Targets in Triple-Negative Breast Cancer

The correlation between the expression of each core target and tumor stage was evaluated using the staging plot function in the GEPIA2 database. The correlation between the expression of each target and patient survival prognosis was examined using kmplot (https://kmplot.com/analysis/). Information on the expression of each core target in normal and tumor samples was extracted from The Cancer Genome Atlas (TCGA) and The Genotype-Tissue Expression (GTEx) databases.

2.12 Molecular Docking Verification of Quercetin with Core Targets

The protein structures of core targets were downloaded from the Protein Data Bank (PDB) database (https://www.rcsb.org/). PubChem (https://pubchem.ncbi.nlm.nih.gov/) was used for the 3D structure of quercetin. The protein structures and the 3D structure of the drug were imported into CB-Dock2 for molecular docking.

# 3. Results

3.1 Clustering Analysis and Annotation of Single-cell Sequencing Data of Triple-negative Breast Cancer

The single-cell RNA sequencing data of TNBC included in this study was obtained from the GSE161529 dataset, which contains 15,870 cells from patients with triple-negative breast cancer. To obtain the marker genes of various cell types, PCA analysis was performed on the differentially expressed genes screened, and the distances between cells were calculated using the first 25 principal components (PCs) after PCA dimensionality reduction. Based on the PCA-reduced space, the neighboring cells of each cell were determined (defaulting to Euclidean distance) to prepare for clustering. The differences are then analyzed, and the top ten differential genes are shown in Figure 1 A. Through UMAP clustering analysis, the cells in the 17 principal components were divided into 7 cell subtypes, namely: epithelial cells, T cells, fibroblasts, vascular endothelial cells, macrophages, pericytes, and plasma cells (as shown in Figure 1B-C). The differentially expressed genes of T cells were exported for further analysis.

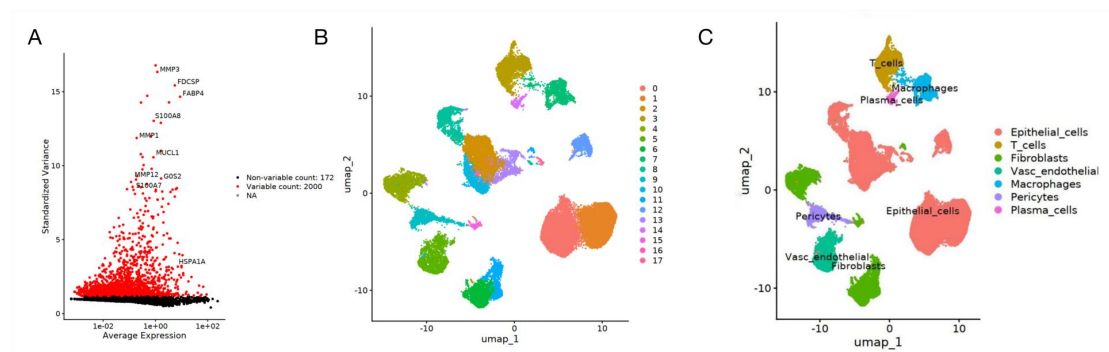

Figure 1 Clustering analysis and annotation results of single-cell sequencing data of triple-negative breast cancer A. PCA analysis results of GSE161529; B. UMAP dimensionality reduction results of GSE161529; C. Cell annotation results of GSE161529.

3.2 Analysis of T-cell Differentially Expressed Genes and Breast Cancer-related Genes

There were a total of 4,847 differentially expressed genes related to T cells in GSE161529. After removing duplicates, there were 7,171 breast cancer-related genes in the CTD, PharmGkb, DisGeNET, and GeneCards databases. The intersection genes included 2,322 genes such as GZMA, CD8A, PRF1, CD2, TIGIT, CD3D, KLRB1, CTLA4, CD7, and CTSW (as shown in Figure 2A). Pathway enrichment analysis was performed on the intersection genes. The KEGG enrichment analysis is shown in Figure 2B, including Epstein-Barr virus infection, Human T-cell leukemia virus 1 infection, Apoptosis, etc. The GO enrichment results are shown in Figure 2C. The enriched pathways in Biological Process (BP) include viral process, ribonucleoprotein complex biogenesis, viral life cycle, etc. The enriched pathways in Cellular Component (CC) include focal adhesion, cell-substrate junction, secretory granule lumen, etc. The enriched pathways of Molecular Function (MF) include cadherin binding, ubiquitin-like protein ligase binding, ubiquitin protein ligase binding, etc.

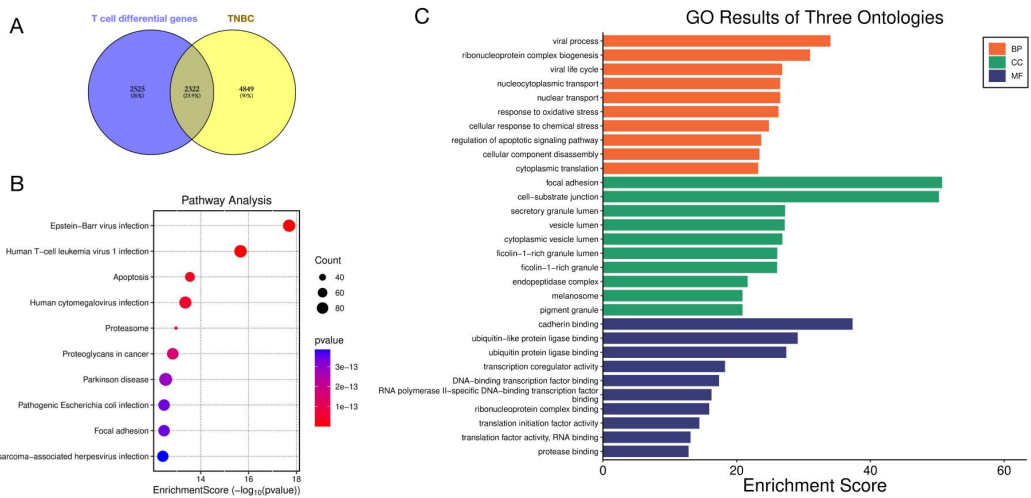

Figure 2 Analysis of T cell differential genes and triple-negative breast cancer-related genes A. Venn diagram of the intersection genes of T cell-related differential genes and breast cancer-related genes in GSE161529; B. KEGG pathway enrichment of the intersection genes of T cell-related differential genes and breast cancer-related genes in GSE161529; C. GO pathway enrichment of the intersection genes of T cell-related differential genes and breast cancer-related genes in GSE161529.

3.3 Analysis of Enriched Pathway-related Genes and Screening of Relevant Traditional Chinese Medicines and Monomers

Search for each pathway in GeneCard, and select the top 50 genes with a score value greater than 60 in each pathway. The KEGG pathway results are shown in Table 1, and the GO pathway results are shown in Table 2. Select the top 10 targets of the relevant pathways, which are aTP53, TNF, EGFR, AKT1, IL6, TGFB1, STAT3, MAPK1, CTNNB1, and STAT1. Input these genes into SymMap and screen for relevant traditional Chinese medicines with the conditions of P_value < 0.05, FDR (BH) < 0.05, and FDR (Bonferroni) < 0.05. The results are shown in Table 3. Search for relevant traditional Chinese medicines in TCMSP with the conditions of OB ≥ 30% and DL ≥ 0.18. A total of 996 compounds were retrieved, and after removing duplicates, 554 remained. The top 10 compounds are shown in Figure 3, which are quercetin, β-sitosterol, kaempferol, luteolin, stigmasterol, stigmasterol, (+)-catechin, isorhamnetin, naringenin, and eupatilin.

| Channel name | Key genes (top 10 shown) |
| --- | --- |
| Epstein–Barr virus infection | MAGT1、STAT3、CFTR、LOC130068460、SH2D1A、TLR3、IFNG、DOCK8、IL10、CCR |
| Human T–cell leukemia virus 1 infection | STAT3、TP53、NF1、RAG1、JAK3、IFNG、CD4、TNF、IL10、ADA |
| Apoptosis | BCL2、CASP3、CASP8、BAX、XIAP、FAS、TP53、TNFSF10、CFLAR、CASP9 |
| Human cytomegalovirus infection | STAT3、DOCK8、RAG1、IFNG、TNF、TLR3、CCR5、IL10、IL6、STAT2 |
| Proteasome | PSMB8、PSMB4、PSMB9、PSMB10、PSMA3、PSMB5、POMP、PSMB1、PSMA5、PSMA1 |
| Proteoglycans in cancer | BRCA2、BRCA1、ATM、TP53、APC、CDH1、EGFR、PTEN、NF1、CDKN2A |
| Parkinson disease | LRRK2、PRKN、PINK1、SNCA、PKHD1、SYNJ1、PKD1、GBA1、PARK7、NPC1 |
| Pathogenic Escherichia coli infection | CFTR、MUTYH、STAT3、DOCK8、APC、TLR3、TNF、IFNG、IL10、IL6 |
| Focal adhesion | ITGB2、PTK2、FERMT3、TSC2、GRIN2A、TSC1、INF2、ICAM1、VCAM1、CNTNAP2 |
| Kaposi sarcoma–associated herpesvirus infection | IL6、STAT3、TLR3、ADA2、BPTF、IFNGR1、IRF3、TP53、IFNG、IL6-AS1 |

Table 1 KEGG pathway-related genes

| Gene ontology class | Channel name | Key genes (top 10 shown) |
| --- | --- | --- |
| Biological Process (BP) | viral process | STAT2、TLR3、STAT1、IFIH1、IFNAR1、CD4、TNF、MIR7-3HG、IFNG、TP5 |
| | ribonucleoprotein complex biogenesis | SDHA、PEX1、PEX5、PEX2、PEX14、PEX19、PRKAR1A、SURF1、NDUFS1、LRPPRC |
| | viral life cycle | TP53、STAT2、TNF、TLR3、STAT1、OTC、CCND1、CDK1、CDKN1A、MIR7-3HG |
| | nucleocytoplas | ABCB1、NUTF2、TAP1、SLC3A2、INS、INSR、 |
| | mic transport | RAN、APOA1、XPO1、ABCD4 |
| | nuclear transport | SDHA、SLC22A5、SLC37A4、FOLR1、SLC6A4、TMEM70、SLC2A1、SLC6A3、LRPPRC、SURF1 |
| | response to oxidative stress | BRCA1、TP53、ATM、RYR1、BRCA2、TNF、NOS3、CFTR、NF1、IL6 |
| | cellular response to chemical stress | TP53、BRCA1、ATM、BRCA2、RYR1、TNF、NF1、CFTR、EGFR、IL6 |
| | regulation of apoptotic signaling pathway | TP53、ATM、STAT3、AKT1、MAPK1、CTNNB1、EGFR、PTEN、PIK3CA、MTOR |
| | cellular component disassembly | RAG1、RAG2、TP53、C4B、EGFR、ATM、AKT1、TNF、TERT、CTNNB1 |
| | cytoplasmic translation | EIF4E、EIF3A、EIF4G1、EIF2S1、PABPC1、EIF3B、EEF1A1、EIF4EBP1、EIF4A1、TP53 |
| Cellular Component, CC | focal adhesion | ITGB2、PTK2、FERMT3、TSC2、GRIN2A、TSC1、INF2、ICAM1、VCAM1、CNTNAP2 |
| | cell-substrate junction | TJP1、LAMC2、COL17A1、ITGB4、JUP、ITGA6、MUSK、PLEC、CDH1、CTNNB1 |
| | secretory granule lumen | SLPI、SPINT2、CHGA、CRISP3、TIA1、G3BP1、GUCY2C、ZG16、SRGN、SELP |
| | vesicle lumen | VAMP2、VAMP7、VAMP8、USO1、VAMP3、VTI1B、SEC22B、NSF、VAMP4、DSC3 |
| | cytoplasmic vesicle lumen | VAMP2、VAMP7、VAMP8、VTI1B、DYNC1H1、VAMP3、USO1、SEC22B、APP、NSF |
| | ficolin-1-rich granule lumen | PRG2、LAMP1、VCP、LAMP2、DDX3X、C1orf35、GMFG、COMMD3、GUSB、PTPRN2 |
| | ficolin-1-rich granule | NBEAL2、PRG2、LAMP1、DDX3X、VCP、LAMP2、GUSB、COMMD3、C1orf35、PTPRN2 |
| | endopeptidase complex | SDHA、PRKAR1A、BCS1L、TMEM70、MME、NDUFS1、NDUFA10、MT-CYB、NDUFAF5、LRPPRC |
| | melanosome | PMEL、RAB27A、MYO5A、GPR143、MLPH、TYRP1、TYR、OCA2、AP3B1、BLOC1S6 |
| | pigment granule | TYR、CEBPE、PRPH2、MITF、TYRP1、OCA2、MC1R、RPE65、SERPINF1、SMARCD2 |
| Molecular Function (MF) | cadherin binding | CDH1、CDH2、CTNNB1、CDH5、GATA1、CDH3、CDH13、CTNNA1、CDH11、CDH17 |
| | ubiquitin-like protein ligase binding | TP53、HADHA、GATA1、HADHB、PRKN、F5、MDM2、APP、HSD17B4、CREBBP |
| | ubiquitin protein | TP53、PRKN、LIG4、HADHA、MDM2、GATA1、 |
| | ligase binding | UBE3A、RPS27A、HADHB、ITCH |
| | transcription coregulator activity | STAT3、STAT1、TP53、MAPK1、JUN、AR、ESR1、EP300、SP1、PPARG |
| | DNA-binding transcription factor binding | F8、F5、IGF1R、IGF1、F9、TP53、GATA1、JUN、F11、STAT3 |
| | RNA polymerase II-specific DNA-binding transcription factor binding | TP53、IGF1R、IGF1、F9、JUN、STAT3、GATA1、POLR2L、TNF、TBP |
| | ribonucleoprotein complex binding | SDHA、PRKAR1A、MBL2、SURF1、TP53、HNRNPU、GATA1、LRPPRC、SDHB、NDUFS1 |
| | translation initiation factor activity | F5、F8、F9、IGF1R、IGF1、F2、F11、F7、EIF4E、F10 |
| | translation factor activity, RNA binding | F5、F8、IGF1R、F9、IGF1、TNF、F2、TP53、F10、CFH |
| | protease binding | MBL2、GATA1、MASP2、MASP1、IGFBP3、SERPINA6、PDHX、ABCB1、TP53、SERPINA7 |

Table 2 GO pathway-related genes

| Related genes | Related Chinese medicine |
|---|---|
| TP53 | *Patrinia scabiosifolia、Scutellaria barbata、Ilex cornuta、Polygonum cuspidatum、Spatholobus suberectus、Schizonepeta tenuifolia、Oroxylum indicum、Sophora tonkinensis、Pyrrosia lingua、Potentilla chinensis、Agrimonia pilosa、Hovenia dulcis* |
| TNF | *Ginkgo biloba、Patrinia scabiosifolia、Lobelia chinensis、Scutellaria barbata、Mentha haplocalyx、Microcos paniculata、Atractylodes lancea、Angelica pubescens、Rubus chingii、Alpinia officinarum、Pueraria lobata、Ilex cornuta、Hypericum perforatum、Choerospondias axillaris、Trigonella foenum-graecum、Citrus reticulata 'Chachi'、Sophora japonica、Astragalus membranaceus、Paederia scandens、Centella asiatica、Lonicera japonica、Sedum aizoon、Geranium wilfordii、Glechoma longituba、Forsythia suspensa、Campsis grandiflora、Lysimachia foenum-graecum、Humulus scandens、Ephedra sinica、Verbena officinalis、Portulaca oleracea、Actinidia chinensis、Eclipta prostrata、Oroxylum indicum、Equisetum hyemale、Ligustrum lucidum、Flemingia philippinensis、Peucedanum praeruptorum、Citrus reticulata、Panax ginseng、* |
| | *Mulberry Root Bark、Mulberry Fruit、Mulberry Leaf、Hippophae rhamnoides、Sophora tonkinensis、Crataegus pinnatifida、Gnaphalium affine、Silybum marianum、Polygonum orientale、Aesculus chinensis、Smilax glabra、Potentilla chinensis、Prunella vulgaris、Agrimonia pilosa、Elsholtzia ciliata、Ginkgo biloba、Zea mays、Daphne genkwa、Coriandrum sativum、Gardenia jasminoides、Aster tataricus、Senecio scandens* |
| EGFR | None |
| AKT1 | *Plantago asiatica、Citrus aurantium、Potentilla discolor、Dioscorea hypoglauca、Ailanthus altissima、Drynaria fortunei、Choerospondias axillaris、Euonymus alatus、Arachis hypogaea、Geranium wilfordii、Ephedra sinica、Ephedra Root、Portulaca oleracea、Smilax glabra、Potentilla chinensis、Prunella vulgaris、Agrimonia pilosa、Epimedium grandiflorum、Daphne genkwa* |
| IL6 | *Ginkgo biloba、Patrinia scabiosifolia、Ilex cornuta、Hedysarum polybotrys、Sophora japonica、Geranium wilfordii、Glechoma longituba、Oroxylum indicum、Morus alba、Hippophae rhamnoides、Sophora tonkinensis、Crataegus pinnatifida、Smilax glabra、Gardenia jasminoides* |
| TGFB1 | *Alpinia officinarum、Ilex cornuta、Arachis hypogaea、Smilax glabra* |
| STAT3 | *Ilex cornuta、Potentilla chinensis* |
| MAPK1 | *Potentilla discolor、Ilex cornuta、Euonymus alatus、Cichorium intybus、Oroxylum indicum、Potentilla chinensis* |
| CTNNB1 | None |
| STAT1 | *Ginkgo biloba、Sophora tonkinensis、Potentilla chinensis* |

Table 3 Gene-related TCM screening

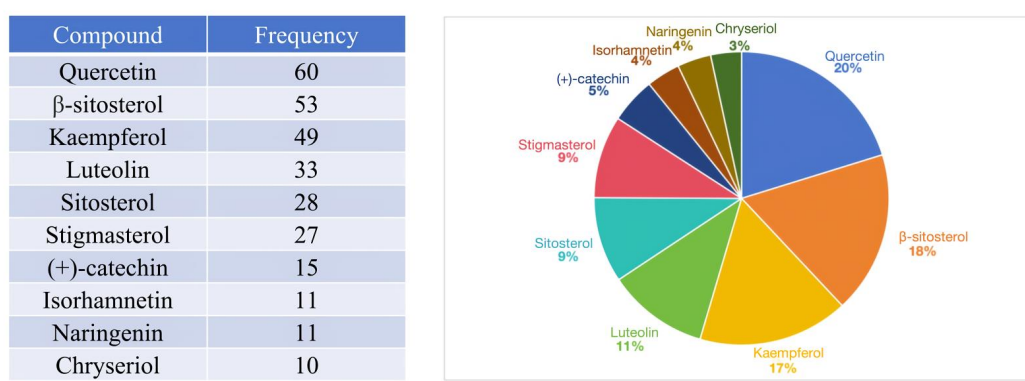

Figure 3 Proportion of relevant TCM monomers

3.4 Quercetin and Breast Cancer-Related Gene Analysis

Quercetin was selected as the research object. Its SMILES number was retrieved from Pubchem and imported into Swiss Target Prediction to predict its target sites. The intersection of quercetin's target sites and those related to triple-negative breast cancer was obtained, resulting in 79 key target sites (as shown in Figure 4A). These include NOX4, AVPR2, AKR1B1, XDH, MAOA, IGF1R, FLT3, CYP19A1, EGFR, F2, CA2, PIM1, ALOX5, AURKB, ADORA1, GLO1, MPO, PIK3R1, DAPK1, PYGL, CA1, GSK3B, SRC, PTK2, KDR, MMP13, MMP3, CA3, ABCC1, PLK1, CDK1, MMP9, CA12, MMP2, PKN1, CA9, CSNK2A1, ALOX12, MET, NEK2, CXCR1, CAMK2B, ALK, AKT1, ABCB1, PLA2G1B, CA5A, BACE1, CYP1B1, AXL, ABCG2, AKR1C2, AKR1C1, AKR1C3, MAPT, TOP2A, INSR, AChE, MYLK, SYK, PIK3CG, APEX1, PTPRS, ESR2, ARG1, CDK6, CDK2, TYR, HSD17B1, AHR, ESRRA, APP, PARP1, TTR, MMP12, CD38, TNKS2, TOP1, and TERT. The KEGG enrichment results of the intersection key targets are shown in Figure 4B, involving Nitrogen metabolism, EGFR tyrosine kinase inhibitor resistance, Endocrine resistance, etc. The intersection targets were imported into the STRING database to obtain the protein-protein interaction network (PPI) of the key targets (as shown in Figure 4C). The data was optimized in Cytoscape software, and the degree values were calculated. Targets with a degree value greater than 20 were considered core targets and were placed in the innermost circle of the concentric circles (as shown in Figure 4D), including AKT1, EGFR, SRC, MMP9, GSK3B, PARP1, KDR, MMP2, CDK2, IGF1R, and MET.

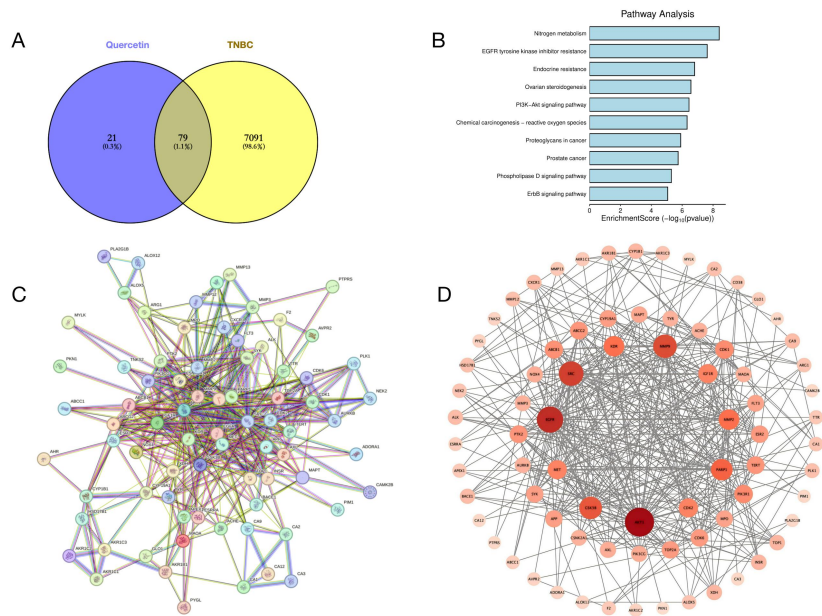

Figure 4 Analysis of Quercetin and Breast Cancer Related Genes A. Venn diagram of the intersection targets of quercetin and triple-negative breast cancer; B. KEGG enrichment results of the intersection targets of quercetin and triple-negative breast cancer; C. Protein-protein interaction map of the intersection targets of quercetin and triple-negative breast cancer; D. Calculation results of interaction Degree values of the intersection targets of quercetin and triple-negative breast cancer.

3.5 Analysis of Quercetin and Core Targets in Breast Cancer

The correlation analysis between each core target and the overall survival of triple-negative breast cancer is shown in Figure 5. Among them, the high expression positively correlated with the overall survival of triple-negative breast cancer are: AKT1, MMP9, and CDK2; the high expression negatively correlated with the overall survival of triple-negative breast cancer are: EGFR, SRC, PARP1, MET, and IGF1R. The expression information of each core target in normal and tumor samples is shown in Figure 6. Significantly increased in tumor samples are: EGFR, KDR, and MET;

significantly decreased in tumor samples are: MMP9 and PARP1. The correlation between the expression of each core target and tumor stage is shown in Figure 7. The results indicate that the expression level of MMP9 significantly changes with cancer stage.

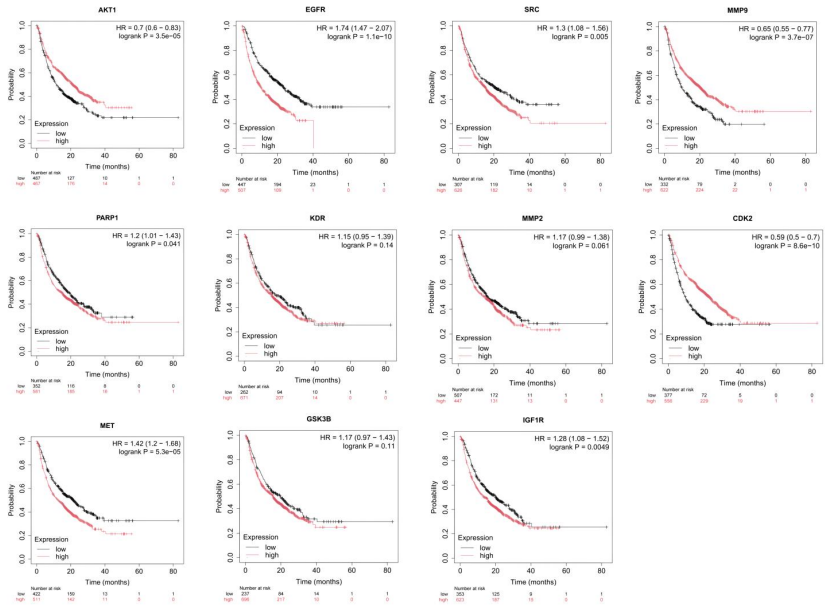

Figure 5 Correlation Analysis of Each Core Target with Overall Survival of Breast Cancer

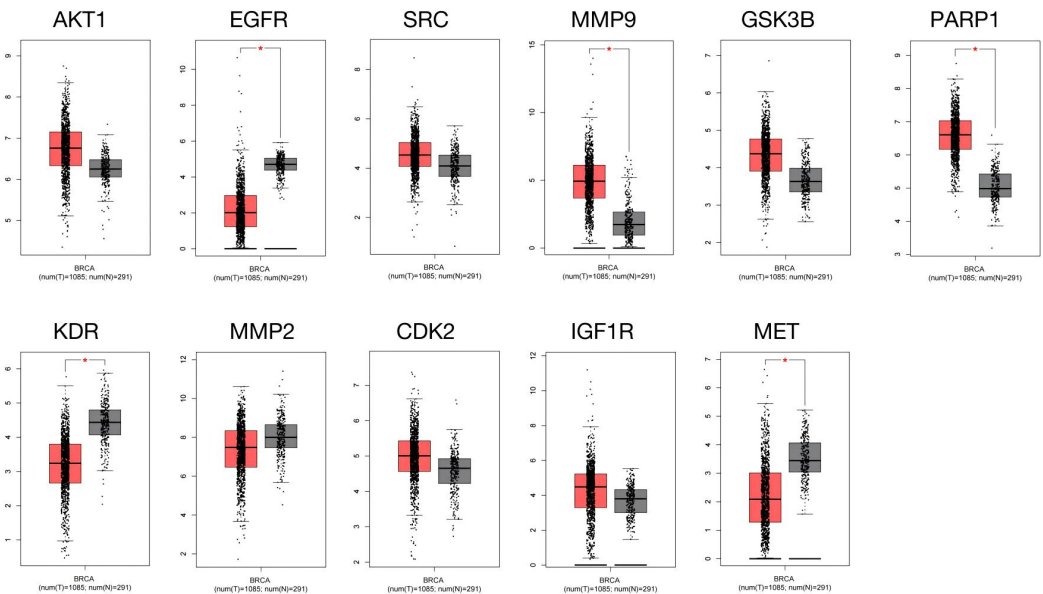

Figure 6 Information on the expression of each core target in normal and tumor

samples

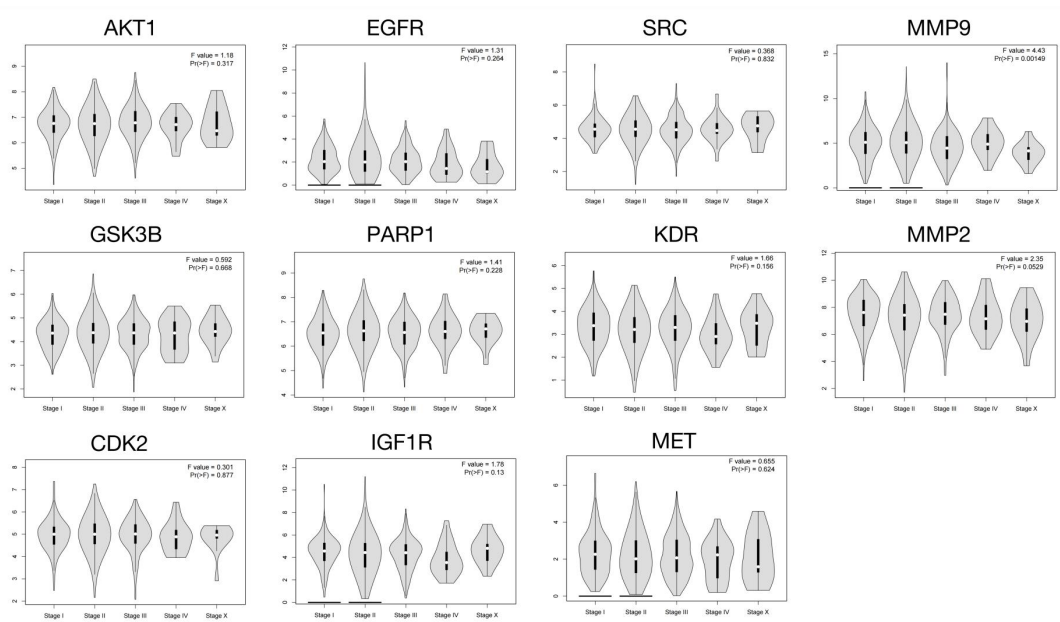

Figure 7 Correlation between the expression of each core target and tumor stage

3.6 Molecular Docking of Quercetin with Core Targets of Breast Cancer

As shown in Figure 8, the docking results of quercetin with each core target were all less than -6 KJ/mol, indicating that the docking of quercetin with each core target was stable. Among them, the docking with GSK3B was the best, with a result of -13.2 KJ/mol.

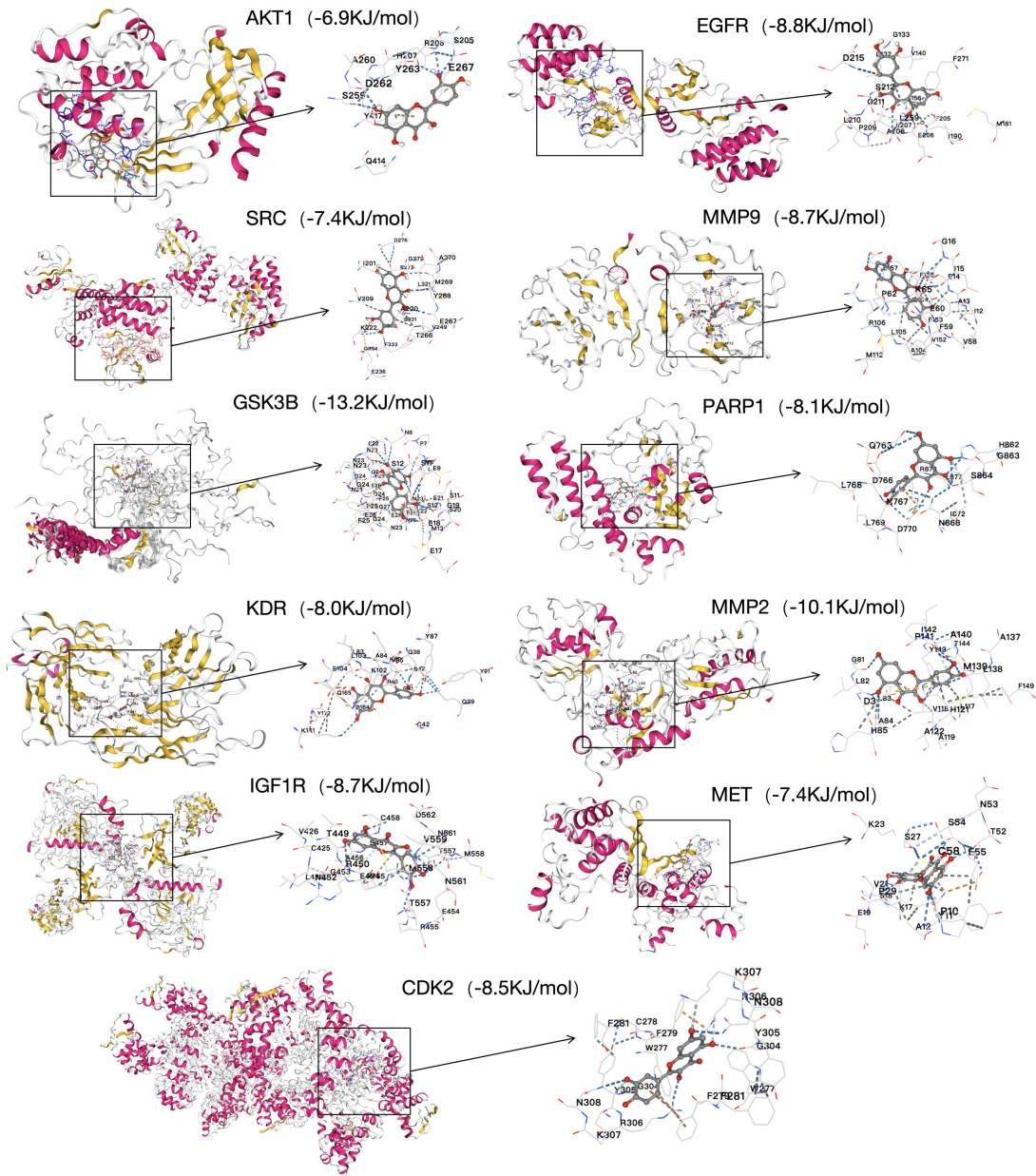

Figure 8 Molecular docking of quercetin with each core target.

# 4. Discussion

This study systematically analyzed the molecular mechanism by which quercetin inhibits triple-negative breast cancer (TNBC) through regulating T cell-related targets by integrating single-cell transcriptome sequencing and network pharmacology methods. The research results not only revealed the key targets of T cell dysfunction in the TNBC immune microenvironment but also verified the potential application value of quercetin as a natural compound in the immunotherapy of TNBC. The following discussion will be conducted from multiple perspectives on the research results.

4.1. Characteristics and functions of T cell subsets in the TNBC immune microenvironment

Single-cell sequencing analysis provided an unprecedented high-resolution perspective for understanding the heterogeneity of T cells in the tumor microenvironment (TME) of TNBC. Through UMAP dimensionality reduction clustering technology, this study successfully identified multiple functionally distinct T cell subsets in the TNBC TME, including CD8+ effector T cells, regulatory T cells (Tregs), and exhausted T cells, etc. Notably, these subsets exhibited significant heterogeneity in spatial distribution and gene expression profiles. Among them, exhausted T cells (characterized by high expression of PD-1, TIM-3, etc.) were significantly enriched in TNBC, which is highly consistent with the previous finding that TNBC has a strong immunosuppressive characteristic. Further through differential gene analysis, we found that multiple key T cell function genes (such as cytotoxic effector molecules GZMA, T cell markers CD8A, perforin PRF1, etc.) were generally downregulated in the TNBC microenvironment. This widespread inhibition pattern of T cell function-related gene expression suggests that TNBC may induce T cells into a deep exhausted state through multiple mechanisms. Particularly noteworthy is that KEGG pathway analysis showed that these downregulated genes were significantly enriched in key immune-related pathways such as T cell receptor signaling and cytotoxic T cell activation, providing a new molecular-level insight into the low response rate of TNBC patients to existing immunotherapies (such as PD-1 inhibitors). These findings not only deepen our understanding of the immune escape mechanism of TNBC but more importantly provide important theoretical basis and potential intervention targets for the development of novel targeted therapeutic strategies (such as dual immune checkpoint blockade therapy targeting CTLA-4 and PD-1, or the development of small molecule agonists that can reverse T cell exhaustion). Future studies can further combine spatial transcriptome technology to explore the precise localization of these T cell subsets in tumor tissues and their spatial interaction with other immune cells, thereby providing a more comprehensive scientific basis for the development of more precise immunotherapy strategies.

4.2. Key Targets and Pathways of Quercetin in Regulating TNBC

This study employed a systems network pharmacology approach to deeply analyze the multi-target mechanism of quercetin in regulating TNBC. By integrating multiple database resources, we identified 79 common target proteins between quercetin and TNBC, which exhibited significant modular characteristics in the protein-protein interaction (PPI) network. Notably, core targets such as AKT1, EGFR, and MMP9 showed high network centrality (Degree > 20), suggesting their potential key regulatory roles in the quercetin-TNBC interaction network. KEGG pathway enrichment analysis revealed that these targets were significantly enriched in multiple signaling pathways closely related to TNBC treatment resistance, particularly the EGFR tyrosine kinase inhibitor resistance pathway and the endocrine resistance pathway, indicating that quercetin may overcome TNBC drug resistance through multi-target synergy. Specifically, AKT1, as a core node of the PI3K/AKT/mTOR signaling axis, its abnormal activation is directly related to TNBC chemotherapy resistance [18] and may also participate in immune escape by regulating PD-L1 expression; while MMP9, as an important member of the matrix metalloproteinase family, not only promotes tumor metastasis but also remodels the immune microenvironment by degrading the extracellular matrix and inhibiting T cell infiltration [19]. Molecular docking studies provided structural biological evidence for these theoretical predictions, showing that quercetin could form stable complexes with all core targets (binding energy < -6 kJ/mol), with the strongest interaction being with GSK3B (binding energy -13.2 kJ/mol). This high-affinity binding may inhibit the kinase activity of GSK3B, thereby affecting key signaling pathways such as Wnt/β-catenin[20]. Notably, survival analysis indicated that the expression levels of AKT1 and MMP9 were significantly correlated with the prognosis of TNBC patients, providing potential efficacy prediction markers for the clinical translation of quercetin. These findings not only reveal the advantages of quercetin as a multi-target drug but also offer new ideas for developing combined treatment strategies for TNBC drug resistance (such as quercetin combined with AKT inhibitors). Future research can further verify the specific contributions of these targets in the anti-tumor effect of quercetin by constructing gene-edited TNBC cell models.

4.3. The Regulatory Potential of Quercetin on T Cell Function

This study innovatively adopted a multi-omics integration analysis strategy, combining single-cell transcriptome sequencing technology with systems network pharmacology methods to comprehensively analyze the multi-target regulatory network of quercetin on T cell function in the TNBC immune microenvironment from single-cell resolution to the system level. The research results showed that quercetin enhanced the anti-tumor activity of T cells through multi-dimensional and multi-level immune regulatory mechanisms: at the level of immune checkpoint regulation, quercetin could specifically target overexpressed co-inhibitory molecules (such as CTLA-4, TIGIT, LAG-3, etc.) in the TNBC microenvironment, by competitively binding to the active sites of these immune checkpoints and blocking their interaction with ligands, thereby reversing the functional exhaustion state of T cells and restoring the cytotoxicity (manifested as upregulated expression of granzyme B and perforin) and proliferation ability (increased Ki-67 positive rate) of CD8+ T cells. In terms of the remodeling of the cytokine network, quercetin exhibits a unique "bidirectional regulatory" effect: on the one hand, it significantly inhibits the production of immunosuppressive factors such as TGF-β1, IL-6, and IL-10; on the other hand, it promotes the secretion of Th1-type cytokines such as IFN-γ, TNF-α, and IL-2. This coordinated effect effectively improves the immunosuppressive characteristics of the tumor microenvironment, creating favorable conditions for the function of T cells[21]. In terms of signal pathway intervention, integrated analysis reveals that the target sites of quercetin are significantly enriched in multiple key pathways related to TNBC treatment resistance, including the EGFR signaling pathway, the PI3K-AKT-mTOR pathway, and the JAK-STAT pathway, etc. Notably, quercetin can simultaneously target and regulate multiple key nodes in these pathways (such as AKT1, STAT3, mTOR, etc.). This multi-target cooperative action mode may help overcome the treatment resistance caused by compensatory activation of pathways commonly seen in TNBC. These important findings not only clarify the mechanism by which quercetin enhances T cell anti-tumor immunity at the molecular level but also provide a scientific basis for the development of new immunotherapy strategies based on natural products.

4.4. Innovation and Limitations of the Research

The innovation of this study mainly lies in two aspects: First, a breakthrough in methodological integration, for the first time combining single-cell RNA sequencing technology with network pharmacology analysis methods to construct a multi-level research framework from single-cell resolution to systems pharmacology, comprehensively analyzing the multi-target action network of quercetin in regulating T cell function in the TNBC microenvironment, providing a new technical route for the study of the mechanism of action of natural products; second, an important breakthrough in target discovery, through bioinformatics mining and molecular docking verification, not only confirming the effect of quercetin on known targets (such as EGFR), but more importantly, identifying core targets such as AKT1 and MMP9 that have not been fully focused on previously, which are closely related to the immune escape and drug resistance of TNBC, providing important clues for the development of new targeted treatment strategies. However, the study also has certain limitations: at the level of mechanism verification, the current conclusions are mainly based on computational biology predictions, and it is urgent to conduct functional verification through in vitro T cell function experiments and TNBC mouse models and other in vivo and in vitro experiments; at the clinical application level, the inherent low water solubility and metabolic instability of quercetin may lead to insufficient bioavailability, and future research should focus on exploring the construction of nano-delivery systems (such as liposomes, polymer nanoparticles) or improving its pharmacokinetic properties through structural modifications (such as glycosylation, methylation) to enhance its clinical translation potential. These innovative discoveries and existing challenges jointly point out the key directions for future research, laying an important foundation for promoting the application of quercetin in TNBC immunotherapy.

4.5. Future Research Directions

Based on the important findings of this study, future research can be further explored in three key directions: First, in terms of mechanism verification, it is necessary to establish a TNBC mouse model, systematically evaluate the regulatory effect of quercetin on the infiltration degree, subpopulation distribution, and functional status (such as cytotoxicity, proliferation ability) of T cells in the tumor microenvironment through flow cytometry, immunohistochemistry, and other techniques, and verify the contribution of key targets through gene knockout experiments; second, in terms of combined treatment strategies, the synergistic effect of quercetin with PD-1/PD-L1 inhibitors should be focused on, optimizing the dosing regimen through dose gradient experiments, and using multi-omics technology to analyze the molecular mechanism by which it enhances the efficacy of immune checkpoint inhibitors; Finally, in view of the pharmacokinetic limitations of quercetin itself, its bioavailability and tumor targeting can be enhanced through structural optimization strategies (such as introducing targeting groups, prodrug design) or the development of novel delivery systems (such as exosome drug loading, stimulus-responsive nanoparticles), laying the foundation for clinical translation. These research directions will jointly promote the transformation of quercetin from basic research to clinical application.

## 5. Conclusion

This study has revealed the molecular mechanism by which quercetin regulates the function of TNBC T cells through multiple targets, providing theoretical support for its application as an adjuvant in immunotherapy. Despite the challenges in translation, with the combination of delivery technologies or combination drug strategies, quercetin holds promise as a new option for the treatment of TNBC.

**Declarations**

**Ethics approval and consent to participate**

Not applicable.

**Consent for publication**

All authors agree to publish this article.

**Data Availability**

Not applicable.

**Funding**

This research was supported by the Guangdong Provincial Administration of Traditional Chinese Medicine (20231217).

**Competing interests**

All authors declare no conflicts of interest.

**Authors' contributions**

Chenruiqi completed the search and organization of data and the writing of the paper, Han Liang and Wang Fengyun supervised and revised the paper.